\documentclass[10pt]{iopart}
\usepackage{iopams}
\usepackage{epsfig}
\newcommand{\be}{\begin{equation}}
\newcommand{\ee}{\end{equation}}

\begin{document}

\title[Principal bifurcation branch]
{On the principal bifurcation branch of a third
order nonlinear long-wave equation}
\author{ R. D. Benguria and M.\ C.\ Depassier\footnote[3]{To
whom correspondence should be addressed.}}
\address{
 Facultad de F\'{\i}sica, Pontificia Universidad Cat\'olica de
Chile, Casilla 306, Santiago 22, Chile}


\begin{abstract}
We study the principal bifurcation curve of a third order equation
which describes the nonlinear evolution of several systems with a
long--wavelength instability. We show that the main bifurcation
branch can be derived from a variational principle. This allows to
obtain a close estimate of the complete branch. In particular,
when the bifurcation is subcritical,  the large amplitude stable
branch can be found in a simple manner.
\end{abstract}


\section{Introduction}
\label{Introduction}

Long wavelength instabilities have been found  in diverse physical
systems, such as  Rayleigh--B\'enard convection between insulating
boundaries \cite{Hurle67}, Marangoni convection in the presence of
a free surface \cite{Scriven64,Smith66,Takashima81}, directional
solidification \cite{Sivashinsky83,Riley90}, Langmuir circulation
in the ocean \cite{Cox94} and others \cite{Oron97}. Evolution
equations that describe the nonlinear evolution of the instability
can be derived by means of an expansion procedure based on the
long scale of the instability \cite{Chapman80}. Unlike the case of
instabilities with non zero wavelength where the weakly nonlinear
evolution of the instability is governed by the Ginzburg--Landau
equation, for long wavelength instabilities different equations
arise according to the problem considered \cite{Nepomnyashchy95}.
In the absence of a unique equation to describe these
instabilities each equation needs to be examined  to determine
whether it provides an adequate description of the problem under
study.  It is often found that the bifurcation from the basic
steady state is subcritical, thus unstable. In such cases it is
important to determine whether large amplitude stable solutions
exist. The usual procedure to capture the turning back of a
subcritical branch is to perform an asymptotic expansion close to
the bifurcation point. Suitable scaling of parameters is necessary
in order to capture stable solutions. If this procedure fails one
may search for approximate large amplitude solutions of the
equation which is not always feasible. The alternative is to
obtain the bifurcation curve numerically.

The purpose of this work is to study a prototype long wave
evolution equation which, in a given parameter range, bifurcates
subcritically from the basic state \cite{DS81,DS82},  and show
that the principal bifurcation branch derives from an integral
variational principle. It is then possible to obtain lower bounds
on the complete branch that are as close as desired to the exact
solution.  While we treat a specific equation, the method can be
applied to other equations as well. In Section II the main result
is derived. In Section III a particular exactly solvable case is
discussed. Analytical and numerical  results for the general case
are given in Section IV.

\section{Formulation of the problem}

We consider the long--wavelength equation \be f_t + \lambda
f_{\xi\xi} + \kappa f_{\xi\xi\xi\xi} - \nu (f_\xi^3)_\xi +
 \mu (f_\xi f_{\xi\xi})_\xi + \delta (f f_\xi)_\xi= 0,
\label{first} \ee in a periodic box of length $2 L$. The boundary
conditions $ f_\xi(\pm L) = f_{\xi\xi\xi}(\pm L) =0$ are chosen to
satisfy physical requirements. With these boundary conditions the
space average $\overline f$ is conserved.

This equation describes the weakly nonlinear evolution of
Rayleigh--B\'enard and Marangoni convection between insulating
boundaries \cite{DS81,DS82,Gertsberg81,Depassier84,Oron89}.
Particular cases of steady state solutions to this equation have
been studied in connection with directional
 solidification.  In this context the case  $\nu=0$ is known as  the
(steady state) Riley--Davis equation with vanishing segregation
coefficient. The case $\nu=0$ and $\mu=0$ is known as the
Sivashinsky equation. The bifurcation structure of these two
equations has been studied extensively \cite{Sivashinsky83,
Riley90, Novick92, Sarocka99}. In the context of
Rayleigh--B\'enard convection, the term proportional to $\mu$
arises due to the asymmetry between upper an lower mechanical
boundary conditions \cite{Chapman80} and the term proportional to
$\delta$ arises when non--Boussinesq effects are included
\cite{DS81,DS82,Depassier84}. The purpose of this work is to study
the  {\it principal} (monotonic) branch  that bifurcates from the
first eigenvalue. These solutions are sometimes called {\it
simple}\cite{Novick92}.

Introducing a new length scale $x = \xi /L$, scaling the amplitude
as $f = \sqrt{\kappa/\nu}\, \phi$, and integrating once, the equation
can be written as
\begin{equation}
r \phi_x + \phi_{xxx} - \phi_x^3 + A \phi_x\phi_{xx} + B \phi
\phi_x =0, \label{maineq}
\end{equation}
for appropriate constants $A$ and $B$. The constant of integration
in the equation above vanishes due to the boundary conditions. Due
to the symmetry $x\rightarrow -x$ we may consider the solution on
half the interval and impose
\begin{equation}
\phi_x(0) = \phi_x(1) =0.
\label{bcs1}
\end{equation}
In addition, for the principal branch we impose
\begin{equation}
\phi(0) = \phi_m, \qquad \qquad \phi(1) =0, \qquad
\qquad \phi_x<0. \label{bcs2}
\end{equation}
 Previous studies of this
equation  show that   the quadratic term $B \phi \phi_x$ is
destabilizing, it gives rise to a subcritical instability
(\cite{Sivashinsky83,DS81,DS82}). If $B<0$ the bifurcation is
supercritical. The bending back of the unstable branch can be
described close to the transition point $B=0$ by means of small
amplitude expansions. Expanding the solution  in a small parameter
$\epsilon$, as $\phi(x) = \epsilon (\phi_0 + \epsilon \phi_1 +
...)$ and the eigenvalue as $r = r_0 +\epsilon r_1 +....,$ we
obtain for the eigenfunction in leading order
\[
\phi_0(x) = \phi_m \frac{1 + \cos[\pi x]}{2},
\]
and at order $\epsilon$
\[
\phi_1(x) = \frac{1}{48\pi^2}(B - A \pi^2)(\cos[2\pi x] - 1).
\]
The eigenvalue up to order $\epsilon^2$ is given by
$$
r = \pi^2 - \epsilon B \frac{\phi_m}{2} + \epsilon^2
\frac{\phi_m^2}{96\pi^2} [18 \pi^4 + (B- \sqrt{2}\pi^2 A)^2
-(3-2\sqrt{2})\pi^2 A B].
$$
For positive $B$ the bifurcation is subcritical, but the order
$\epsilon^2$ term shows that it regains stability for small enough
$B$. This result is valid only for small amplitude and
sufficiently low values of $B$.

Our purpose is to show that the principal branch $r(\phi_m)$ can
be obtained from a variational principle from which lower bounds
on the bifurcation curve are obtained. This enables us to obtain
non perturbative results valid  at arbitrary large amplitude and
for all parameter values. The method used to derive this result is
related to that developed to study a class of reaction diffusion
problems \cite{BD96a,BD96b,BD04} and others \cite{BD96c}.

The first step is to rewrite the equation in phase space, therefore we define
\[
p(\phi) = - \frac{d \phi}{d x}
\]
and write the problem to be solved as \cite{DS82}
\be r p + p
\frac{d }{d\phi}\left( p \frac{d p}{d\phi}\right) - p^3 + A p^2
\frac{d p}{d\phi} + B\phi p = 0, \label{eqp} \ee with \be p(0) =
p(\phi_m) =0, \qquad \qquad p>0. \label{bcps} \ee Even though the
equation above can be reduced to quadratures, an explicit
expression for the eigenvalue $r(\phi_m)$ cannot be obtained
except in the case $B=0$ (Appendix A). From the variational
principle that we construct below, the exact solution  for the
case
 $B=0$ can be recovered, and arbitrarily accurate bounds can be obtained when  $B\neq 0$.

Let $g(\phi)$ be an arbitrary positive continuous  function, with
continuous second derivatives, such that $g(0) = g(\phi_m) =0$.
Multiplying Eq.(\ref{eqp}) by $g/p$ and integrating by parts we
obtain, after regrouping terms, \be r \int_0^{\phi_m} g(\phi)
\,d\phi + B \int_0^{\phi_m} \phi\, g(\phi) \,d\phi =
\int_0^{\phi_m} p^2(\phi) h(\phi) \,d\phi, \label{arreglo} \ee
where we have defined \be h(\phi) = -\frac{1}{2} g'' + g +
\frac{A}{2}g'. \label{hache} \ee

Boundary terms which appear when integrating by parts vanish
because of (\ref{bcps}) and the conditions on $g$. In addition we
require that the function $g$ be such that $h>0$ in $(0,\phi_m)$,
a condition which can always be met (see Appendix B).

We now use H\"older's inequality,
\[
\left( \int_0^{\phi_m} |s(\phi)|^\ell\,d\phi \right)^{1/\ell}
\times \left( \int_0^{\phi_m} |q(\phi)|^k\,d\phi \right)^{1/k}\ge
\int_0^{\phi_m} s(\phi) q(\phi)\,d\phi, \] with $$\frac{1}{\ell} +
\frac{1}{k} =1
$$
to bound the right side of equality (\ref{arreglo}). Choosing $s=
(p^2 h)^{1/3}$, $q = p^{-2/3}$, $\ell=3$ and $k=3/2$ we obtain
\[
\left( \int_0^{\phi_m} p^2 h \,d\phi \right)^{1/3} \left(\int_0^{\phi_m} \frac{1}{p} d\phi
\right)^{2/3} \ge \int_0^{\phi_m} h^{1/3} d\phi.
\]
Using
\[
1 = \int_0^1 d x = \int_0^{\phi_m} \frac{d\phi}{p},
\]
the above inequality becomes \be \int_0^{\phi_m} p^2 h \,d\phi \ge
\left[\int_0^{\phi_m} h^{1/3} d\phi \right]^3. \label{ineq} \ee
Inserting (\ref{ineq}) in (\ref{arreglo}) we obtain the desired
lower bound, \be r \ge \frac{ \left[\int_0^{\phi_m}
h^{1/3}(\phi)\, d\phi \right]^3- B \int_0^{\phi_m}
 \phi g(\phi) \,d\phi }{\int_0^{\phi_m} g(\phi) \,d\phi}.
\label{main1}
\ee
To show that this is a variational principle we must prove that there exists
 a function $g=\hat g$ for which equality holds in (10) \cite{BD96a}.
  In H\"older's inequality, the case of equality corresponds to  $s^\ell = q^k$.
  In the present case we have that equality in (\ref{ineq}) holds when
\be h = \hat h = \frac{1}{p^3}. \ee From the definition of $h$ we
have equality in (\ref{main1}) when $g$ is a solution of
\begin{equation}
-\frac{1}{2}  g'' +  g + \frac{A}{2} g'= \frac{1}{p^3},
\label{eq:11b}
\end{equation}
satisfying $g(0) = g(\phi_m) =0$.  To prove that the solution to
this boundary value problem, which we call $\hat g$,  exists and
that it yields finite values for the appropriate integrals in
(\ref{main1})
 is straightforward (see Appendix B for details).

Since equality holds for a special function $g$ we conclude that the principal bifurcation
curve of (\ref{maineq}) is given by
 \be
r = \max_g \frac{ \left[\int_0^{\phi_m} h^{1/3}(\phi) \,d\phi \right]^3- B \int_0^{\phi_m}
 \phi g(\phi) \,d\phi }{\int_0^{\phi_m} g(\phi)\, d\phi},
\label{main2} \ee where the maximum is taken among all positive
continuous functions $g$ (with continuous second derivatives and
such that $g(0)=g(\phi_m)=0$)  for which $h>0$ and integrals
exist. The maximum is achieved for $g = \hat g$ up to a
multiplicative constant.

It is worth pointing out that the Euler--Lagrange equation for
(\ref{main2})
 can be obtained easily (Appendix C) and is, effectively, equation (\ref{eqp})
 when $g = \hat g$.

Expressions (\ref{main1}) and (\ref{main2}) constitute our main result.
Using adequate trial functions $g$ in (\ref{main1}) we obtain lower bounds on
the bifurcation branch $r(\phi_m)$. Moreover we know there exists a trial function
for which equality can be attained.

In the following section we study the case $B=0$ and show that the
maximizing $\hat g$ can be constructed explicitly in this case.

\section{The case B=0}

In this section we show that the exact formula for the eigenvalue
can be obtained from the variational principle when $B=0$.
Furthermore, the optimal $g$ obtained for this case will provide a
good trial function for $B \neq 0$.

To obtain the solution for the eigenvalue, instead of solving the maximizing problem
(\ref{main2}),  we address the equivalent problem of minimizing its inverse. Let
\be
J_m = \min_g J[g] = \min_g \frac
{\int_0^{\phi_m} g(\phi)\, d\phi}
{\left( \int_0^{\phi_m} h^{1/3}(\phi) \,d\phi\right)^{3}},
\label{jota}
\ee
with $g>0$, and $h >0$ as defined in (\ref{hache}). Given $h$, $g$ can be found in
terms of $h$ by means of the Green's function $G(\phi,\phi')$ for (\ref{hache}). Defining
\be
s(\phi) = h^{1/3}(\phi)
\ee
we have that
\be
g(\phi) = \int_0^{\phi_m} G(\phi,\phi') s^3(\phi') \,d\phi',
\ee
where the Green's function $G(\phi,\phi')$ corresponding to equation (\ref{hache})
 is given by
\begin{equation}
G(\phi,\phi') = \left\{
                \begin{array}{cc}
\frac{2}{b \sinh(b\phi_m)}
\mbox{e}^{A(\phi-\phi')/2}
\sinh( b\phi) \sinh(b(\phi_m-\phi')) &\mbox{for $\phi < \phi'$}\\
\frac{2}{b \sinh(b\phi_m)} \mbox{e}^{A(\phi-\phi')/2}\sinh( b\phi') \sinh(b(\phi_m-\phi))
&\mbox{for $\phi >  \phi'$,}
\end{array}
\right.
\label{eq:15b}
\end{equation}
with
\[
b = \frac{1}{2} \sqrt{A^2 +8}.
\]
Then,
\be
\int_0^{\phi_m} g(\phi) \,d\phi = \int_0^{\phi_m} d\phi'F(\phi') s^3(\phi'),
\ee
where
\begin{eqnarray}
\fl F(\phi) = \int_0^{\phi_m} G(\xi, \phi) d\xi = 1 -
 \mbox{e}^{-A\phi/2 } \nonumber \\
 \lo{\times} \left[ \cosh (b \phi) - \coth(b\phi_m) \sinh(b \phi) +
 \mbox{e}^{A \phi_m/2} \mbox{cosech}(b\phi_m) \sinh(b\phi) \right].
\label{EFE}
\end{eqnarray}
Finally,  (\ref{jota}) can be written as, \be J_m = \min_s J[g] =
\min_s \frac {\int_0^{\phi_m} F(\phi) s^3(\phi) d\phi} {\left(
\int_0^{\phi_m} s(\phi) d\phi\right)^{3}}, \label{jota2} \ee
which is a simple variational problem for which it is
straightforward to find the minimizing function. The minimizing
$s$ is given by
 \be s_{\mbox{\footnotesize{min}}}(\phi) = \frac{1}{N}
\frac{1}{\sqrt{F(\phi)}}, \ee where \be N= \int_0^{\phi_m}
\frac{d\phi}{\sqrt{F(\phi)}}. \label{ene} \ee Using this function
in (\ref{jota2}), we find
\[
J_{\mbox{\footnotesize{min}}} = J(s_{\mbox{\footnotesize{min}}}) =
\frac{1}{N^2}
\]
and the exact value of $r$ for $B=0$ is $1/J(
s_{\mbox{\footnotesize{min}}})$. Spelled out explicitly,
\[
r(\phi_m) = \left(\int_0^{\phi_m} \frac{d\phi}{\sqrt{F(\phi)}}\right)^2,
\]
which is the exact  expression for the bifurcating branch when B=0
(Appendix A). Notice that we have also obtained implicitly the
solution to the equation itself, since equality in (\ref{main2})
holds when $\hat h =  s_{\mbox{\footnotesize{min}}}^3 = 1/p^3$.
The solution of the equation is obtained by integrating $p =-
d\phi/dx = 1/s_{\mbox{\footnotesize{min}}}. $

\section{Bounds for $B\neq 0$}

The exact solution for the principal branch cannot be obtained in
closed form when $B\neq 0$. Instead, upper bounds can be obtained
using different trial functions in (\ref{main1}). In this section
accurate bounds will be obtained numerically. A simple analytical
bound will be obtained in the special case $A=0$.

\subsection{Numerical bounds}

 Guided by the results of the previous section,  we choose as a  trial
function  the exact solution for the case  B=0, that is,
\begin{equation}
g_1(\phi) = \int_0^{\phi_m} G(\phi,\phi')
s_{\mbox{\footnotesize{min}}}^3(\phi') d\phi'. \label{try1}
\end{equation}
Replacing this choice for $g$ in (\ref{main1}) we obtain \be r \ge
\left(\int_0^{\phi_m} \frac{d\phi}{\sqrt{F(\phi)}}\right)^2-
\frac{B}{N} \int_0^{\phi_m}
\frac{1}{F^{3/2}(\phi')}\left(\int_0^{\phi_m} \phi G(\phi,\phi')
d\phi\right)d\phi' \ee where $N$ is  defined in (\ref{ene}).

A different, simpler trial function, is obtained  choosing
$h(\phi)$= 1, which gives
\begin{equation}
g_2(\phi) = \int_0^{\phi_m} G(\phi,\phi') d\phi'. \label{try2}
\end{equation}

In Figure 1 below we show the bifurcation curve for A=0, B=5. The
solid line is the exact numerical solution obtained using the
software Auto \cite{Doedel}. The dot--dashed line is the
 lower bound obtained using the trial function (\ref{try1}), and the dashed line is the
 bound obtained with the trial function (\ref{try2}). The bound obtained using $g_1$ as a trial
 function  provides a very close estimation of
 the bifurcating branch, especially at low amplitudes. Already the simple trial
 function $g_2$ which does not give a close estimate is enough to prove that the
  subcritical bifurcation is stabilized at larger amplitudes.
\begin{figure}[h]
\begin{center}
\epsfig{file=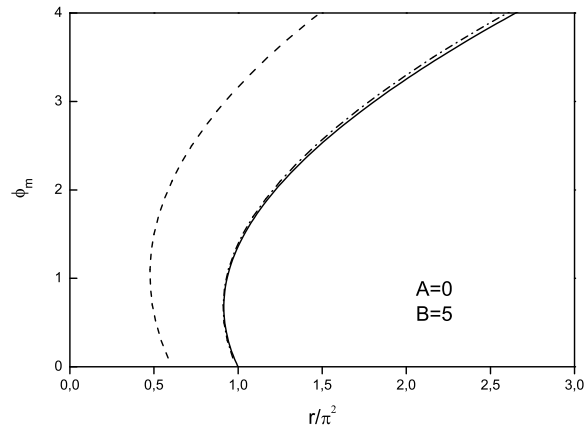,height=7.0truecm}
 \caption{Exact
bifurcation curve and bounds with different trial functions.}
\end{center}
\end{figure}
In Figure 2 we show the numerical results for A=0, B=10. The
curves shown are as in Figure 1. At small amplitudes $g_1$ gives a
close estimate, at large amplitudes it captures the qualitative
behavior, still providing a good bound. Here too the simple trial
function $g_2$ suffices to capture the stabilization of
 the subcritical branch at larger amplitudes.

In Figure 3 we show the numerical results for A=10, B=10. The
solid line is the numerical solution of the bifurcation problem
using Auto, the dashed line is the bound obtained with $g_1$.
\begin{figure}[h]
\begin{center}
\epsfig{file=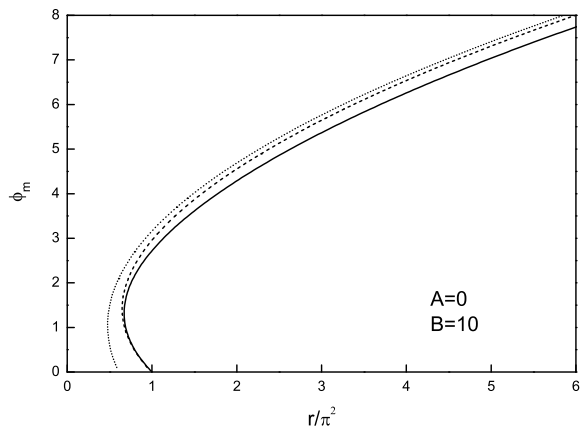,height=7.0truecm} \caption{As in Fig.1
for different parameter values.}
\end{center}
\end{figure}
\begin{figure}[h]
\begin{center}
\epsfig{file=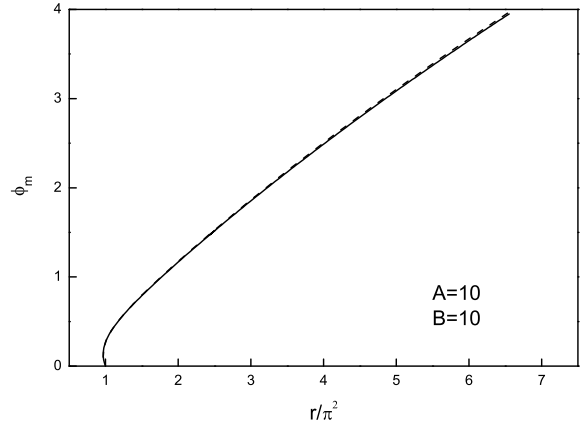,height=7.0truecm} \caption{Exact
bifurcation curve and bound with the B=0 exact solution}
\end{center}
\end{figure}

\subsection{Analytical bounds}

In the case A=0 a simple admissible trial function is $g_3(\phi) =
\phi (\phi_m- \phi)$. Then,
\[
h(\phi) = 1 + \phi (\phi_m - \phi) > \phi (\phi_m - \phi).
\]
Using $g_3(\phi)$ and this lower bound for the corresponding
$h(\phi)$ in (\ref{main1}) we obtain
\begin{eqnarray}
r &\ge& \frac{ \int_0^{\phi_m} [\phi(\phi_m-\phi)]^{1/3}\,d\phi -
B \int_0^{\phi_m} \phi^2 (\phi_m-\phi)\,d\phi} {\int_0^{\phi_m}
\phi (\phi_m-\phi)\,d\phi} \nonumber \\
&\ge& 6 B(4/3,4/3) \phi_m^2 - \frac{B}{2} \phi_m \nonumber \\
&\ge& 3,18 \phi_m^2 - \frac{B}{2} \phi_m, \label{solan}
\end{eqnarray}
where $B(x,y)$ is the standard Beta function. This bound, valid
for all amplitudes and all values of $B$ shows that the
subcritical branch always turns around and becomes stable at
sufficiently large amplitude.

\subsection{Bounds for solutions of zero average}

Here we address the problem of solutions to (\ref{maineq}) of zero
average. The bifurcating branch $r(\phi_m)$ obtained above does
not correspond to  a solution of fixed average. Each point along
the branch has different average, $\overline \phi = \overline\phi
(\phi_m)$. It is possible to obtain bounds for solutions of zero
average as follows. Define
 $u = \phi - \overline \phi$. In terms of $u$, equation
 (\ref{maineq}) becomes
 \[
 (r + B\overline\phi) u_x + u_{xxx} - u_x^3 + A u_x u_{xx} + B u
 u_x =0,
 \]
 with
 \[
 u_x(0) = u_x(1) =0, \qquad u(0) = \phi_m- \overline\phi, \quad
 u(1) = -\overline\phi.
 \]
 The amplitude of the function $u$ is $\phi_m$, its average is
 zero and the bifurcating branch $r_{\overline u=0}$ is given by
 \[
 r_{\overline u=0}(\phi_m)= r(\phi_m) + B \overline \phi.
 \]
Since $0 \le \overline \phi \le \phi_m$, we know that for $B>0$,
$B \overline \phi >0$, whereas for $B<0$, $B \overline \phi >
B\phi_m$. Then we know that for a solution of zero average
\begin{equation}
r_{\overline u=0} \ge \left\{ \begin{array}{cc}
                            r(\phi_m) & \mbox{ for $B>0$} \\
                            r(\phi_m)+ B \phi_m & \mbox{ for $B<0$}.
\end{array}
\right.
\end{equation}
The analytical bound (\ref{solan}) for $A=0$  becomes for
solutions of zero average,
\begin{equation}
r_{\overline u=0} \ge \left\{ \begin{array}{cc}
                            3.18 \phi_m^2 - B\phi_m/2 &\mbox{ for $B>0$} \\
                            3.18 \phi_m^2 + B\phi_m/2 & \mbox{ for
                            $B<0$}.
\end{array}
\right.
\end{equation}
This shows that also for zero average solutions, all subcritical
branches turn back at sufficiently large amplitude.

\section{Summary}

We have studied the principal bifurcation branch of a third order
equation which exhibits a subcritical bifurcation for certain
parameter values. The bifurcation problem can be characterized by
an integral variational principle from which lower bounds on the
complete branch can be found. With the use of appropriate trial
functions it is possible to obtain an accurate estimation of the
complete branch,  without resource to perturbation theory. This
permits to capture the stabilization of the subcritical branch
even when the turning point occurs at large amplitude.

While we focused on a specific equation, the method can be applied
to other nonlinear equations [18-21].

\ack This work was partially supported by projects Fondecyt
1020844 and 1020851.

\appendix
\section{}

In this appendix we show that even though the problem the problem
can be reduced to quadratures, an explicit expression for the
eigenvalue can be obtained only when $B=0$.

Equation (\ref{eqp}) can be reduced to a linear equation for $u=
p^2$, \be \frac{d^2 u}{d \phi^2} + A \frac{d u}{d \phi} - 2 u + 2
(r+ B\phi) =0. \ee The solution to this equation which satisfies
the boundary conditions $u(0) = u(\phi_m) =0$ is given by \be u =
(r + \frac{A B}{2}) F(\phi) + B Z(\phi), \ee with $F(\phi)$ as
defined in (\ref{EFE}) and \be Z(\phi) = \phi - \phi_m
\mbox{e}^{-A(\phi-\phi_m)/2} \mbox{cosech}(b \phi_m) \sinh(
b\phi). \ee The eigenvalue $r$ follows from the condition \be
\int_0^{\phi_m} \frac{d \phi}{p(\phi)}=1, \ee which may be written
as \be \int_0^{\phi_m} \frac{d\phi}{\sqrt{(r + AB/2)F(\phi) + B
Z(\phi)}} = 1. \ee Since this integral cannot be performed
analytically, only in the case $B=0$ an explicit expression for
$r$ is found.

\section{}

In this appendix we analyse the properties of the solution of
(\ref{eq:11b}), i.e., the solution of the two point boundary value
problem
\begin{equation}
-\frac{1}{2} g'' + \frac{A}{2} g' + g = \frac{1}{p^3}
\label{eq:B1}
\end{equation}
with $g(0)=g(\phi_m)=0$, where $p(\phi)$ is a solution of (\ref{eqp})
with $p(0)=p(\phi_m)=0$. One can conveniently write (\ref{eq:B1}) as
\begin{equation}
L(g) = \frac{2}{p^3} e^{-A \, \phi},
\label{eq:B2}
\end{equation}
where the linear operator $L$ is given by
\begin{equation}
L = - \frac{d}{d \phi} \left( e^{-A\phi} \frac{d}{d \phi} \right) + e^{-A\phi}.
\label{eq:B3}
\end{equation}
The operator $L$ acting on functions $g$ that vanish in $0$ and $\phi_m$
is self--adjoint (with respect to the usual inner product
$(f,g) =\int_0^{\phi_m} f(\phi) g(\phi) \, d \phi$) and positive definite
(to be precise $L$ is acting on $H^1_0(0,\phi_m)$). In fact the Green's
function associated to the operator $L$ (i.e., the kernel of $L^{-1}$) is
explicitly given by $e^{A \phi'}G(\phi,\phi')$ where $G$ is given by
(\ref{eq:15b}). Notice that $e^{A \phi'}G(\phi,\phi')$ is symmetric under
the exchange of $\phi$ and $\phi'$, and pointwise positive in the interval
$(0,\phi_m)$.

Since the right side of (\ref{eq:B2}) is nonegative for the
solutions to (\ref{maineq}) that we are considering, i.e., for the
{\it principal branch}, and since $L$ is positive definite, the
corresponding solution $g$ to the boundary value problem
(\ref{eq:B1}) is positive in $(0,\phi_m)$. By standard regularity
properties of differential equations, the solution $g$ is
continuous in $(0,\phi_m)$. In the neighborhood of the end points,
the solution $g$ of (\ref{eq:B1}) behaves as follows: $g(\phi)$
vanishes like $\sqrt{\phi}$ near $0$, whereas $g(\phi)$ vanishes
like $\sqrt{\phi_m - \phi}$ near $\phi_m$. To determine this
behavior, notice that the solutions of (\ref{maineq})
corresponding to the {\it principal branch} satisfying the
boundary conditions (\ref{bcs1}), (\ref{bcs2}) are real analytic.
Thus, in the neighborhood of $x=0$ (because of the boundary
conditions), $\phi(x) \approx \phi_m - c \, x^2$, for some $c >0$,
hence, $p=- \phi'(x) \approx 2 c x$ near $x=0$, and inverting
(i.e., expressing $x$ in terms of $\phi$ we finally have $p(\phi)
\approx \sqrt{(\phi_m-\phi)}$ in the neighborhood of
$\phi=\phi_m$. Analogously, $p(\phi) \approx \sqrt{\phi}$ in the
neighborhood of $\phi=0$. From the properties of the Green's
funtion for the operator $L$ (notice that $G$ vanishes linearly
near each of the endpoints) we see that the solution of
(\ref{eq:B1}) also behave as $\sqrt{\phi}$ near $\phi=0$ and as
$\sqrt{(\phi_m-\phi)}$ near $\phi=\phi_m$.

Concerning the existence of the different integrals that appear on
the right side of (\ref{main1}) when $g$ is the solution of
(\ref{eq:B1}), notice that $h$ is continuous in $(0,\phi_m)$, and
although $h$ is singular near the end points, $h^{1/3}$ behaves
like $1/\sqrt{\phi}$ (respectively like $1/\sqrt{\phi_m - \phi}$)
in the neighborhood of $0$ (respectively in the neighborhood of
$\phi_m$), so $h^{1/3}$ is locally integrable near the end points.
The other two integrals also exist, because $g$ is continuous in
$[0,\phi_m]$.

\section{}

In this appendix we derive the Euler-Lagrange equation for
(\ref{main2}). Consider the maximization of the functional \be
K[g] = \left[\int_0^{\phi_m} h^{1/3}(\phi) \,d\phi \right]^3- B
\int_0^{\phi_m} \phi g(\phi) \,d\phi, \ee where \be h(\phi) =
-\frac{1}{2} g'' + g + \frac{A}{2}g'. \ee subject to
\[\int_0^{\phi_m} g(\phi) d\phi =1.\] Introducing the constraint
with a Lagrange multiplier $\lambda$, we obtain the Euler-Lagrange
equation \be \left(\int_0^{\phi_m} h^{1/3}d\phi\right)^2
\left[\frac{1}{h^{2/3}} - \frac{A}{2} \frac{d}{d
y}(\frac{1}{h^{2/3}}) -\frac{1}{2} \frac{d^2}{d
y^2}(\frac{1}{h^{2/3}})\right] - B \phi + \lambda =0. \ee At the
extrema $h= 1/p^3$ and the above equation becomes \be \left(
\int_0^{\phi_m} \frac{d\phi}{p}\right)^2 \left[ p^2 - \frac{A}{2}
\frac{d p^2}{d y} - \frac{1}{2} \frac{d^2 p^2}{d y^2}\right] - B
\phi + \lambda =0. \ee Identifying the Lagrange multiplier
$\lambda$ with the eigenvalue $r$ and observing that  the integral
in the equation above is equal to 1, we verify that  the Euler
Lagrange equation for extrema of $K[g]$ is the original equation
(\ref{eqp}) whose solution we seek.
\end{document}